# Site-Specific Plan-view (S)TEM Sample Preparation from Thin Films using a Dual-Beam FIB-SEM

Supriya Ghosh[1], Fengdeng Liu[1,2], Sreejith Nair[1], Bharat Jalan[1], K. Andre Mkhoyan[1]

1. *Department of Chemical Engineering and Materials Science, University of Minnesota, Minneapolis, MN 55455 USA.*
2. *Department of Electrical and Computer Engineering, University of Minnesota, Minneapolis, MN 55455 USA.*

## Abstract

Plan-view transmission electron microscopy (TEM) samples are key to understand the atomic structure and associated properties of materials along their growth orientation, especially for thin films that are stain-engineered onto different substrates for property tuning. In this work, we present a method to prepare high-quality plan-view samples for analytical STEM study from thin-films using a dual-beam focused ion beam scanning electron microscope (FIB-SEM) system. The samples were prepared from thin films of perovskite oxides and metal oxides ranging from 20-80 nm thicknesses, grown on different substrates using molecular beam epitaxy. A site-specific sample preparation from the area of interest is described, which includes sample attachment and thinning techniques to minimize damage to the final TEM samples. While optimized for the thin film-like geometry, this method can be extended to other site-specific plan-view samples from bulk materials. Aberration-corrected scanning (S)TEM was used to access the quality of the thin film in each sample. This enabled direct visualization of line defects in perovskite $BaSnO_3$ and Ir particle formation and texturing in $IrO_2$ films.

## Introduction

Focused Ion Beam (FIB) is one of the most commonly used instruments in preparation of cross-section samples from bulk materials for transmission electron microscopy (TEM) studies, due to its widespread applicability to many different material systems including ceramics, soft materials like polymers and biological specimens, as well as nano-scale electronic and magnetic devices.[1–4]

In most dual-beam FIB-SEM systems, simultaneous monitoring of the TEM lamella during the preparation can be achieved using the electron- and ion-beam imaging, allowing precise monitoring of the sample quality. This allows preparation of samples from extremely small (a couple of microns to a few tens of nanometers) areas of interest such as nano-scale devices, grain boundaries and patterns in the material. The Ga ion beam in the FIB can be used to mill material away from the sample to create thin cross-sections into the material (< 50 nm), enabled by the presence of micro-manipulator systems to lift the samples out from bulk, and gas injection systems (GIS) for material deposition during the process.

For TEM studies, optimal sample thickness is governed by the mean free path of the elastic and inelastic scattering of the probe electrons in the material at the operating conditions (typically in the range of 60-300 keV) and, therefore, is crucial to control.[5] It should be noted that, while thinner samples are desirable, there are challenges in terms of sample bending and curtaining during the thinning stages in the FIB. Amorphization of the material during thinning in the Ga ion beam cannot be avoided and needs consideration as well. Studies have been carried out in the optimization of the cross-section TEM sample preparation using the FIB to address these issues, with focus on minimizing material damage during the thinning with careful adjustments of the ion-milling conditions.[6,7] Advances in FIB technology, such as availability of cryo-FIBSs to work on beam-sensitive specimen as well as alternate ion-sources such as Xe or plasma FIBs provide further help.[8–11] However, there are many cases where plan-view samples from specific areas are needed. They are essential for the investigation of film textures, grain sizes and orientations, and defects along the film growth direction, and much more.

While methods like chemical etching[12], ion-milling[13] and tripod polishing have been used to obtain electron transparent plan-view samples, preparation of samples from specific location of the material in study still remains very challenging. Some plan-view TEM sample preparation methods have been previously discussed using a FIB, wherein either the samples is loaded perpendicularly for cross-sectioning parallel to the film surface or includes series of complex maneuvering steps for sample preparation using techniques like encapsulation.[14–17]This becomes much harder to implement for thin film samples, which have a film growing on a substrate that needs to be carefully removed. This necessitates the development of a FIB-based technique for plan-view sample preparations, applicable to a wide range of materials and systems.

In this report, we present a new method for preparing plan-view TEM samples from specific sites of interest using a dual-beam FIB-SEM instrument. The method is described and discussed on thin films grown on substrates by molecular beam epitaxy (MBE) with thicknesses ranging from 20 to 75 nm. Three major steps and considerations in the STEM sample preparation have been discussed with the associated instrument settings. The samples prepared were tested using aberration-corrected scanning transmission electron microcopy (STEM) imaging and compositional analysis, to answer key structural aspects for the materials under investigation. Two thin-film material systems are considered here: (i) perovskite oxide $BaSnO_3$ and (ii) metal oxide $IrO_2$ for defect and film texture analysis, correspondingly.

## Methods

Hybrid molecular beam epitaxy (MBE) was used to grow thin films of $BaSnO_3$ and $IrO_2$. For the growth of perovskite $BaSnO_3$ on $SrTiO_3$ (001) substrate, a conventional effusion cell for barium, hexamethylditin (HMDT) as a metal-organic precursor for tin, and an inductively coupled radio frequency (RF) plasma for oxygen was used.[18] Films were grown at a fixed substrate temperature of 950 °C. The substrates were cleaned at 950°C *in-situ* with oxygen plasma prior to growth. Ba was sublimed from a titanium crucible separately with its beam equivalent pressure (BEP) measured by a retractable beam flux monitor before growth. The oxygen flow was set to 0.7 standard cubic centimeters per minute (sccm) to achieve an oxygen background pressure of $5 \times 10^{-6}$ Torr while applying 250 watts of RF power to the plasma coil. HMDT vapor entered the chamber through a heated gas injector (E-Science, Inc.) in an effusion cell port that was in direct line-of-site to the substrate. The HMDT bubbler was held at ~75 °C to provide sufficient HMDT vapor pressure.

For growth of $IrO_2$ on $TiO_2$ (001), the substrate was cleaned sequentially in acetone, methanol and isopropanol before being subjected to a baking process at 200 °C for 2 h in a load lock chamber. All substrates were annealed in oxygen plasma for 20 min at growth temperature before film growth. Ir was supplied by sublimation of 99.9% pure $Ir(acac)_3$ (American Elements), an air-stable solid metal-organic powder, which was placed in a crucible (E-Science) inside a custom-built low-temperature effusion cell (E-Science). An effusion cell temperature of 175 °C was used for Ir supply for all samples. A radio frequency plasma source (Mantis) with charge deflection plates, operated at a forward power of 250 W, was used for supplying reactive oxygen species required for Ir oxidation. An oxygen pressure ~5 × 10$^{-6}$ Torr, supplied using a mass flow controller (MKS Instruments), was used for all film growths. All films were cooled to 120 °C after growth in the presence of oxygen plasma, to avoid formation of oxygen vacancies and surface decomposition of $IrO_2$ to Ir metal. [19]

The cross-section and plan-view sample preparation were carried out on an FEI Helios G4 Dual Beam focused ion beam scanning electron microscope (FIB-SEM) system. The thin film samples for $BaSnO_3$ on $SrTiO_3$ and $IrO_2$ on $TiO_2$ were sputter coated with 50 nm of amorphous carbon (am-C) prior to loading in the FIB. FIB Cu lift out grids (from Ted Pella) were used for sample attachment. The Cu grids were loaded in a TEM row bar from FEI for the Helios system. The FIB sample preparation was carried out using a 30 keV Ga ion beam followed by lower beam energies of 2-5 keV for final thinning steps. For the trenching steps, regular Si rectangular cross-section patterns were utilized. For thinning of the samples, cleaning cross-section patterns were utilized.[20] For SEM imaging during the sample preparation stages, a 15 keV beam energy with a probe current of 0.1 nA was used and a 2-5 keV beam was used in the final thinning stages to get better surface contrast.

Scanning transmission electron microscopy (STEM) measurements were carried out on an aberration corrected FEI Titan G2 (S)TEM 60-300 microscope which is equipped with a CEOS-DCOR probe corrector, a monochromator and a SuperX energy dispersive X-ray (EDX) detector. The microscope was operated at 200 keV with probe currents of 120-140 pA for high angle annular dark-field (HAADF)-STEM imaging and EDX acquisitions. The probe convergence angle was 18.2 mrad with the HAADF inner and outer collection angles of 55 and 200 mrad respectively.

## Results and Discussions

Thin film of $BaSnO_3$ and $IrO_2$ (10-100 nm thick) grown on strain-matched substrates by MBE are used for plan-view TEM sample preparation. To study the atomic structure and properties of these films along the growth direction (or plan-view), the thin film needs to be free of the substrate. Hence, at locations of interest, the substrate needs to be removed from the film using the FIB. Since the Ga ion beam of the FIB can damage the film surfaces and lead to implantation of Ga into the material, the samples are protected by coating the film surface with am-C using a sputter deposition system. A 50-100 nm thick layer of am-C was deposited, prior to loading in the dual-beam FIB-SEM system. The benefits of this coating will be realized during the later thinning stages, wherein the layer can act as an amorphous support for the substrate free film. The steps to create plan-view TEM samples from the area of interest in the sample are described by grouping them into three major steps. These steps with the associated instrument settings and parameters are summarized in Table 1. The settings listed here were optimized for $BaSnO_3$ films but can be slightly adjusted to work on other materials by accounting for the sputtering rates on interaction with the Ga ion beam.

**Table 1:** TEM Lamella preparation steps with corresponding FIB settings and parameters

| Step | Process | Ion Beam Current (nA) | Stage Tilt (°) | Stage Rotation (°) | Pattern Type | Pattern Dimensions [L×W×D] (µm) | Additional Comments |
|---|---|---|---|---|---|---|---|
| 1.1 | Region Identifier | 0.001 | 52 | 0 | Si Line/Circle | 15×0.01<br>1×0.01 | Depth chosen based on film thickness, marked for thinning |
| 1.2 | Protective C \| Pt | 0.09<br>0.26 | 52 | 0 | C/Pt dep | (10-20) × (6-10) ×0.5<br>(10-20) × (6-10) × (1.5-2) | Two-step deposition, for less Ga induced damage |
| 1.3 | Trench 1: Bottom Wedge | 9.1 | 25 | 0 | Regular CS | [(10-20) ± 5] ×15×40 | 5 µm extra length on both sides to provide room for detachment without redeposition; depth (D) chosen to create wedge |
| 1.4 | Trench 2: Top Wedge | 9.1 | 25 | 180 | Regular CS | [(10-20) ± 5] ×15×40 | Create other side of the wedge |
| 1.5 | Trench 3: Side | 9.1 | 52 | 0 | Cleaning CS | 8×10×40 | Detach sample from bulk, with thin bridge left over |
| .2.1 | Bottom Detachment | 1.2 | 0 | 0 | Rectangular | [(10-20) ± 5] ×1.5×5* | *Pattern run till all material is removed |
| 2.2 | Easy Lift Needle Attachment | 0.026-0.041 | 0 | 0 | Pt Weld | 2×2×1 | Needle positioned in both electron and ion beams |
| 2.3 | Lamella Detachment | 1.2 | 0 | 0 | Regular \| Cleaning CS | 10×2×5* | Mill till the sample is free; Cleaning CS on left edge to make uniform |
| 3.1 | Cu Grid Cleaning | 20-65 | 0 | 0 | Cleaning CS | 20×2×40 | Prepare grid for attachment |
| 3.2 | Sample Attachment | 0.041-0.09 | 0 | 0 | Pt Weld | 15×4×4 | Pattern with overlap on grid and sample to fill gap |
| 3.3 | Easy Lift Detachment | 1.2 | 0 | 0 | Rectangular | 4×4×4* | Draw on edge, monitor live for material removal |
| 4.1 | Protective C | 0.001 0.09 0.26 | 52 | 0 | C/Pt dep | 15×2×0.05 15×2×0.5 15×2×1.5 | Protective layer deposition on film region prior to thinning |
| 4.2 | Sample Thinning | 0.001-0.07 | 52 | 0/180 | - | 15×w*×d* | See Table 2 for details. *Size chosen based on sample dimensions |

*Step 1: TEM Lamella Preparation*

First, the area of interest to be studied is chosen by surveying the film surface (Figure 1(a)). An example could be in regions containing grain boundaries, or atomic structure of the film in the regions used for device measurements, or areas after mechanical testing in indentation experiments. Once located, "identifiers" are created to mark the film surface, by using the patterning function of the Ga ion beam with a 1 pA probe as shown in Figure 1(b), to create minimum damage to the film. The depth of these patterns is chosen based on the thickness of the film, as it can be identified during the later thinning steps. The typical dimensions of the samples prepared ranged from 10-25 µm in length and 8-12 µm width, which are much larger than the typical areas used for cross-section sample preparation. Larger areas can also be selected, but would increase the overall sample preparation time. Additional protective layers of am-C or Pt of ~ 1-2 µm were deposited in the area of interest (Figure 1(c)) using the GIS in the FIB. The thicker protective layers ensure the film is protected from Ga ion implantation during the subsequent high current trenching steps to remove the bulk material from the neighboring areas. The ion-beam deposition is done at a stage tilt of 52°, so that the ion beam is perpendicular to the film surface. Next, trenches are created around the sample to free it from the bulk material for detachment. For this, the "regular cross-section" pattern around the protected area is utilized, with the dimensions, much larger than the sample size to ensure easy detachment.

For TEM cross-section samples, the trenching steps are carried out at a stage tilt of 52°, so that the ion-beam is perpendicular to the surface as it removes the material. However, the width of the cross-section samples is usually 1 µm which is much smaller than the dimensions used in plan-view samples. If the trenching is done under these conditions, it would result in a very thick lamella, with a larger volume of the substrate at the bottom to be removed later to free it from the bulk. This makes the detachment challenging with possibility of material redeposition, ultimately leading to failure. To have a thinner substrate, a steeper incidence angle of the ion beam to the film surface should be utilized. The trenching steps are carried out at a stage tilt of 25°, which results in the formation of thin wedge like structure into the material (Figure 1(d)). Higher beam currents of ~ 9 nA are utilized to ensure less material redeposition as well as faster milling times. The milling depth for these patterns is selected to be ~3-4 times the sample length for the substrate to be thin as seen in Figure 1(f). For other materials, this needs to be adjusted based on the material sputtering rate in the Ga beam. Once the sample is trenched on one side, the stage is rotated by 180° to trench out the other side of the sample as seen in Figure 1(e). Next, the sample is freed from the bulk at the sides using rectangular sections as shown in Figure 1(f), where a small bridge is left behind on one side to stabilize the sample for attachment to the manipulator.

*Step 2: Sample Detachment*

After these steps, it is important to ensure the sample is free from the bulk substrate. This is critical to be able to get the sample out of the trenched area using the manipulator for the subsequent steps. To ensure this, an additional milling step is carried out at a stage tilt of 0°, so that the bottom wedge is visible in the ion beam as shown in Figure 2(a). A rectangular milling pattern is used at the bottom with beam currents of 0.7-1.2 nA as shown in Figure 2 (a). Live contrast variations in the patterning area is monitored until all intensity appears dark as seen in Figure 2(b), indicating the material has been completely removed and the lamella is free from the bottom. The cuts on the walls at the bottom in the electron beam SEM image can also be used to monitor for detachment. Next, the EasyLift manipulator system in the FIB is used to lift the sample out. The needle is carefully positioned onto to the sample using both the beams, as shown in Figure 2(c), wherein it is positioned to be at the top-left edge of the sample in the electron beam and bottom right position in the ion beam. The needle positioning is important to ensure easy detachment after attaching to the TEM grid in the next step. The Pt GIS is then used to weld it to the sample top surface as shown in Figure 2 (c). Next, the last bridge connecting the sample to the bulk material is milled away (Figure 2(d)). Prior to lifting out the needle from the trenched area, it is critical to ensure the sample is free using the ion-beam imaging. In cases, where there is redeposition during the final milling step, the rectangular milling sections should be run again in the bottom and the sides, to free the sample. Then, the sample is slowly lifted out from the trench and a final "cleaning cross-section" pattern is run on the left free end to make it uniform, so that it makes good contact when attaching to the TEM grid in the next step.

*Step 3: Sample Attachment to TEM Grid*

The plan-view sample can be attached to any standard FIB TEM grids, which in our case is the Cu lift out grids, shown in Figure 3(a). The Cu TEM grids are usually loaded onto the sample holder vertically in the microscope. This is done to ensure the typical cross-section samples are parallel to the ion-beam during the thinning steps at 52°. Therefore, this cannot be used to thin the samples along the plan-view orientation, as thinning parallel to the film surface is needed. For correct orientation of sample attachment, the TEM grids should be loaded horizontally into the FIB. Figure 3(a) shows the Cu "M" grids that have been loaded onto the FIB TEM grid holder, which is attached horizontally onto the stage using carbon tape. As a result, the TEM grid is now horizontally oriented in the microscope. Further, this allows easy removal of the grid without damaging it for the later thinning steps. An alternate method could use rotation of the EasyLift manipulator system by 90°, if the functionality is available. Figure 3(b) shows the orientation of the grid "M" posts at 0° stage tilt ready for sample attachment. The sample is attached to one of the edges of the "M" post, by utilizing a side welding approach. Prior to attachment, the grid edge surface is made smooth by using a cleaning cross-section as seen in Figure 3(c). This ensures the sample has good contact with the grid wall and attachment is secure in later steps. Next, the EasyLift needle with the sample is inserted (Figure 3(c)) and, carefully positioned near the grid surface for welding. The ion-beam view is first used to bring down the sample near the grid in the z-direction. The electron beam view is then used to slowly position the sample in the x-y directions. As the sample approaches the grid, the shadow of the sample on the grid wall can be used to aid the final maneuvering steps to ensure good contact as shown in Figure 3(d). For the final steps, the slowest manipulator speeds ~ 100nm/s are utilized. Finally, Pt is deposited using the GIS system, to weld the sample onto the Cu grid and fill in any gap left between the sample and the grid. During Pt deposition, the electron beam is frequently used to monitor and check for adequate welding as seen in Figure 3(e). The needle can then be cut free from the sample and retracted, with the sample ready for thinning. Finally, the TEM grid holder is loaded back onto the stage in the standard vertical geometry similar to that of a regular TEM cross-section loading. This ensures the film surface is parallel to the ion-beam for thinning.

*Step 4: Sample Thinning for STEM*

Since the film of interest is embedded between the protective am-C on top and the substrate at the bottom, thinning needs to be done carefully to ensure that the substrate is completely removed and the film is not lost. To understand the sample geometry prior to thinning, imaging in the electron and ion-beam are utilized. Figure 4(a) shows the orientation of the ion-beam during the thinning of the sample. Figure 4(b) shows the electron-beam images from the side-view, wherein the different layers of the sample are indicated, with the protective am-C on top and the wedged substrate at the bottom. The corresponding top (protective am-C) and bottom (substrate) view can be monitored in the ion-beam. For all thinning, the sample is oriented such that the film is parallel to the ion beam, by tilting the stage to 52°, wherein the top-down orientations are visualized in the electron beam SEM image (typically operated at 15 keV) and side cross-section orientation in the ion beam image.

First, a protective am-C layer is deposited onto the sample cross-section where the film is located as shown in Figure 4(c). This is done with sequentially increasing current steps, listed in Table 1 (3), as it prevents extensive damage to the film during subsequent thinning. By knowing the thickness of the protective am-C deposited on top in *Step 1*, the film position is estimated and the area for the am-C deposition in the side-view is selected as shown in Figure 4(c). Next, most of the bulk substrate at the bottom of the film is removed by using a 0.45 nA Ga beam current at 30 keV (Figure 4(d)) till the sample is ~ 2 µm thick. The sample from the bottom (substrate-side) is continuously monitored in the electron beam during the thinning stages. Then, the stage is rotated by 180° to view it from the top (protective am-C side) and remove the protective am-C deposited on top of the film in *Step 1*. The stage rotations ensure the right orientation for monitoring in the electron beam. This is done till the overall sample thickness is ~ 1-1.2 µm. Once we have removed most of the bulk substrate and the bulk protective am-C from top, finer thinning is carried out using the 'rectangular cleaning' cross-sections by gradually reducing the beam current with sample thickness as listed in Table 2, resulting in the sample as shown in Figure 4(e). The cleaning cross-sections patterns are ideal for thinning as they produce the least deposition of the milled away material, enabling even thinning of the entire section.

Next, when the sample is about 300 nm thick, or we start seeing features corresponding to the markers created on the film surface in *Step 1*, the ion beam energy is reduced to 2-5 keV and the beam currents are lowered to prevent damage to the film. The markers made in *Step 1* serve as a guide to exercise caution during the final thinning. Further, a lower energy electron beam (5 keV) is used to monitor the changes in the film contrast as shown in Figure 4(f), as the lower beam energy is more sensitive to surface features.[21] Contrast detected in the electron-beam imaging in this step is critical to determine when all the substrate has been successfully removed from the film at the bottom. The contrast essentially arises from the compositional differences between the film and the substrates. As the final substrate is removed, a bright contrast can be seen (Figure 4(f)) which indicates there are thin areas of the film. This is supported on the am-C layer, which acts like a support analogous to the carbon film supported TEM grid. Further ± 2-3° stage tilts can be used to enable thinning of larger areas and effectively remove more of the substrate. An example of such a thinned sample is shown in Figure 4(g), where we can see large areas of bright contrast, corresponding to the area of interest for STEM imaging. Smaller sections or windows of the sample are thinned sequentially to prevent it from bending as seen in Figure 4(g). The thickness of the samples obtained can vary between 10-50 nm. Such thinned sample should be loaded into the STEM holder with the "substrate" side facing the incident STEM probe. This ensures the sub-Å STEM probe interacts with the "film" first, enabling better image resolution.[22] However, the thickness of the left-over am-C should be < 20 nm to minimize additional beam broadening after propagating though crystalline film.

**Table 2:** FIB-SEM settings for rectangular cleaning cross section during lamella thinning.

| Sample Thickness | Ion Beam Energy (kV) | Probe Current (pA) | Stage Tilt (°) | SEM Beam Energy (kV) |
|---|---|---|---|---|
| > 2 µm | 30 | 400-750 | 52 | 15 |
| ~ 2 µm | 30 | 260 | 52 | 15 |
| ~ 1 µm | 30 | 260 | 52 | 15 |
| ~ 600 nm | 30 | 90 | 52 | 15 |
| ~ 300 nm | 30 | 41 | 52 | 15 |
| ~100 nm | 30 | 7-26 | 52 | 5 |
| ~50-60 nm | 5 | 20-60 | 52±2 | 5 |
| < 50 nm | 2 | 44 | 52±2 | 5 |

To demonstrate the plan-view sample preparation strategy, the results from $BaSnO_3$ (50 and 75 nm) and $IrO_2$ (25 nm) are presented. For each film, several samples are studied for investigating material properties and are not included here as they are beyond the scope of this report. In each case, atomic resolution imaging as well as energy dispersive X-ray spectroscopy (EDX) was carried out to characterize the final thin film quality.

The first system studied was BaSnO$_3$ thin films grown by MBE on SrTiO$_3$ substrates, which are used in electronic device applications for its favorable properties like high room temperature conductivities and electron mobilities in doped films.[23–25] The thin film BaSnO$_3$ are known to have several metallic-like defects in their structure such as dislocations and Ruddlesden-Popper faults propagating along the film growth direction.[26–30] Atomic-resolution STEM analysis is needed to characterize them, which requires having plan-view samples.[27]

Figure 5(a) shows the SEM image from the surface of one such BaSnO$_3$ film, where the grain boundaries can be seen uniformly across the entire film. For plan-view STEM sample preparation, the film was sputter coated with 50 nm of am-C to protect the film surface, as discussed above. In cross-section, as can be seen in the STEM image (Figure 5(b)), the film appears to grow uniformly on the substrate with the dislocations seen as bright contrast lines starting from the substrate interface all the way till the film surface. A 16 × 8 μm sized area on the film surface was chosen to prepare the plan-view sample as shown in Figure 4. The sample was thinned from the substrate side till the region of the am-C was visible, as shown in the low-magnification HAADF image, in Figure 5 (c). The final thinning to remove the substrate was performed with a 5 keV Ga ion beam using cleaning cross section patterns, and the process was monitored using the SEM electron beam imaging during the pattern progression. The thickness of the BaSnO$_3$ layer in this plan-view prepared sample was ~ 30 nm, supported on a thin layer of am-C, with increasing thickness as we move away from the Ga beam direction. The am-C film provided additional stability to keep the film along the (001) zone-axis for imaging, not affecting the atomic resolution characterization of the films. The grains seen in the SEM images are clearly visible in this HAADF-STEM image. STEM-EDX elemental maps obtained from a region from this BaSnO$_3$ film (Figure 5(d)) further confirms the film is indeed mostly freed from the substrate. Area where SrTiO$_3$ substrate is still in place can be seen in the bottom of the image. The Ga concentration in the sample, evaluated from the EDX maps appears to be very low, suggesting minimal ion-beam damage to the film.

To further evaluate the quality of these plan-view samples, atomic-resolution ADF-STEM images were obtained (Figure 6 (a)). The dislocations and other defects create local strain-field variations which can be captured through STEM imaging, as seem in the low-angle ADF-STEM image. The atomic resolution imaging as seen in Figure 6(b) from one such area clearly captures the two major types of dislocations in BaSnO$_3$ films: single ([001]/ (100)) edge dislocations (Figure 6(c)) and dissociated ([001]/ (110)) edge dislocations (Figure 6 (d)). The atomic structure of the dislocations seen here is comparable to or better than previous results from thin film samples prepared by polishing techniques,[27] indicating similar or better quality of sample.

Plan-view samples were also prepared from IrO$_2$ films grown on rutile TiO$_2$ substrates. By changing the film thicknesses, different films surface textures with formation of metallic Ir can be obtained due to unique strain relaxation mechanisms in these films.[31,32] Hence, studying the atomic structure of the film in plan-view provides insights into the structure of the IrO$_2$ films. Plan-view samples were prepared from a 26 nm thick IrO$_2$ thin film grown on (001) TiO$_2$ substrate which appears to relax by forming cracks on the substrate as seen in Figure 7(a). Since the film thickness is less than 30 nm, to obtain large free areas of the IrO$_2$ film, the final thinning steps with a 5 keV Ga ion beam were carried out in much slower steps using beam currents between 7-60 pA for extra caution. The back-scatter detector was used in the SEM to monitor for contrast between the Ir and Ti layers, during thinning, to determine areas of bright contrast free from substrate. Since these films are more susceptible to bending in the final thinning stages due to the cracked structure as seen in Figure 7(a), a thicker layer of the protective am-C layer was left behind for support.

Figure 7 (a) shows a plan-view prepared sample from the IrO$_2$ film, where the cracked nature is visible similar to the surface image seen in the SEM image. To study the surface structure of the film, the very thin edge of the sample was studied. An example of such a thinned area is shown in Figure 7(b), where a periodic like contrast seen is from the formation of Moiré contrast due to the lattice mismatch between the film and the substrate. The areas without these Moiré contrasts can be considered free of the TiO$_2$ substrate, and when imaged in HAADF at lower magnifications, show a periodic-like texture in the thin areas of the film. STEM-EDX mapping (Figure 7(c)) from these regions revealed that they were free from the rutile substrates and the features did not arise from Ga beam- induced damage during the thinning stages. High magnification STEM imaging was used to compare the atomic structure of the film, in the bulk thick region (Figure 7(d)), where the (001) IrO$_2$ structure can be confirmed, also seen from the Fast Fourier transform (FFT). However, the atomic structure of the film as seen in Figure 7(e) closer to the top surface reveals a network of sub 5 nm polycrystalline Ir (111) islands, which is confirmed by the ring observed in the FFT. By changing the defocus during image acquisitions in these regions, the visibility of the nanocrystals can be increased (Figure 7(f)). STEM-EDX mapping from these islands shown in Figure 7(f) indicate they are indeed metallic Ir, and have less O signal intensity in these regions. Thus, we were able to demonstrate how the plan view samples enabled the detection of Ir segregation in the thin films, which can be harder to detect by other bulk characterization techniques at these length scales. These results were possible due to the ability to thin down the film close to the surface for better visibility.

## Conclusions

In this work, we describe a step-by-step method to prepare high-quality plan-view samples from specific locations of thin-film samples, with minimum damage to the atomic structure of the films using a FIB-SEM dual beam system. Four main sample preparation stages are presented, including: (1) creation of "identifier" in the region of interest on the film and deposition of protective layers to preserve the film surface; (2) optimized trenching to create wedges at the bottom of the sample for easy detachment from bulk; (3) horizontal attachment to the TEM grid; and (4) the final sample thinning stages. The sequential thinning steps applied to remove the protective layers and the substrates, were adapted from cross-section TEM sample preparation. Monitoring the sample for contrast variation using the electron beam during final steps of substrate removal are critical to preserve the film, especially for thicknesses < 50 nm. The parameters in each of these individual steps can be tuned further based on the material system being used. Applicability of this method was demonstrated by first preparation of plan-view TEM samples for the site-specific atomic-resolution STEM study of line defects in $BaSnO_3$ thin films. Similarly prepared STEM samples from epitaxially strained $IrO_2$ films enabled identification and characterization of sub-5 nm Ir nanoparticle segregation in these films, not capture by other bulk characterization techniques.

## Acknowledgements

This work was supported primarily by the National Science Foundation through the University of Minnesota MRSEC under award numbers DMR-2011401. Film growth was supported by the U.S. Department of Energy (DOE) through Grant No. DE-SC0020211. Parts of this work were carried out in the Characterization Facility, University of Minnesota, which receives partial support from the NSF through the MRSEC and the NNCI (award number ECCS-2025124) programs. The authors thank Dr Nick Seaton for assistance with the FIB. S.G. acknowledges support from a Doctoral Dissertation Fellowship received from the Graduate School at the University of Minnesota.

**Figures:**

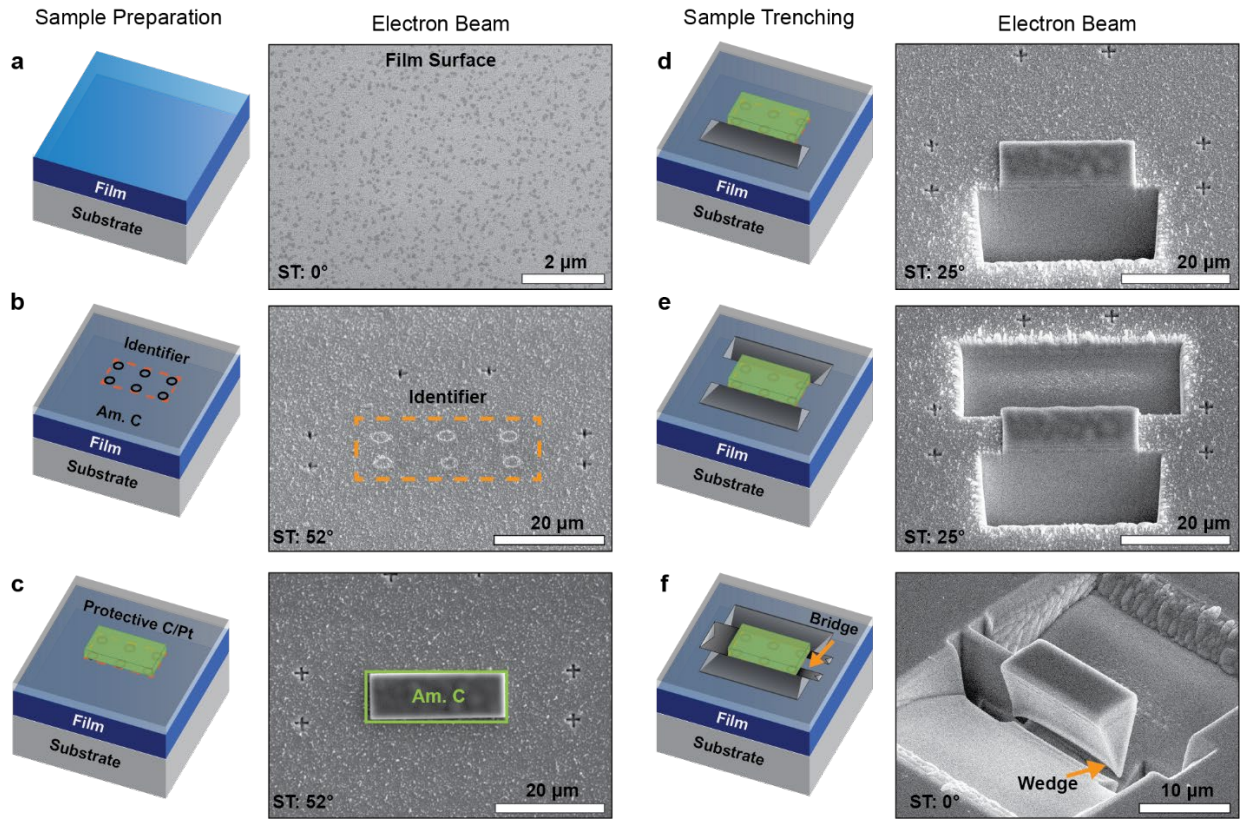

**Figure 1:** TEM Lamella preparation steps. (a) SEM image from the surface of thin-film of BaSnO$_3$ grown on top of SrTiO$_3$ at a stage tilt (ST) of 0 °. (b) ~ 50 nm am-C coated on thin film prior to FIB sample preparation. Identifiers laid down onto the film surface for detection during thinning stages as highlighted within the orange box. (c) ~ 2 µm Protective am-C deposited on marked area prior to trenching to be used for lift-out. (d) Trench 1 cut at the bottom of the sample at a stage tilt of 25°. (e) Stage is rotated by 180° to create identical trench on the other size of the sample (Trench 2). (f) Side-trenches (Trench 3) are created to free material from bulk, with small bridging section left.

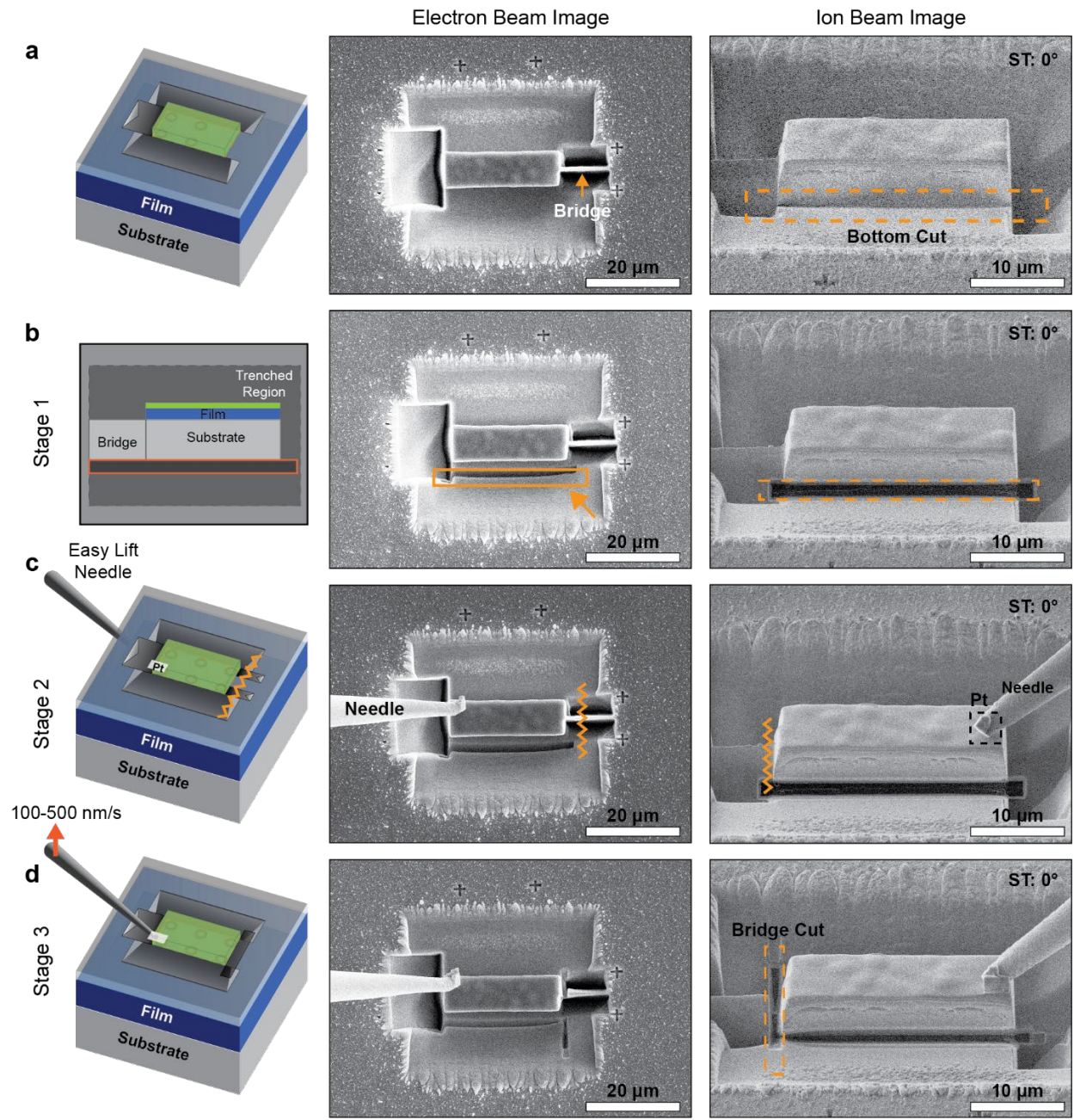

**Figure 2**: Detachment Steps. (a) Bridged sample with area for bottom detachment marked by the orange box in the ion-beam image. (b) Stage 1: Wedge bottom is milled till a dark contrast is observed in the ion beam image and cuts on bulk wall are seen in the electron-beam image. (c) Stage 2: The easy lift needle carefully positioned on sample surface and welded using Pt GIS. (d) Stage 3: Connecting bridge to bulk is cut free and the needle is slowly lifted out.

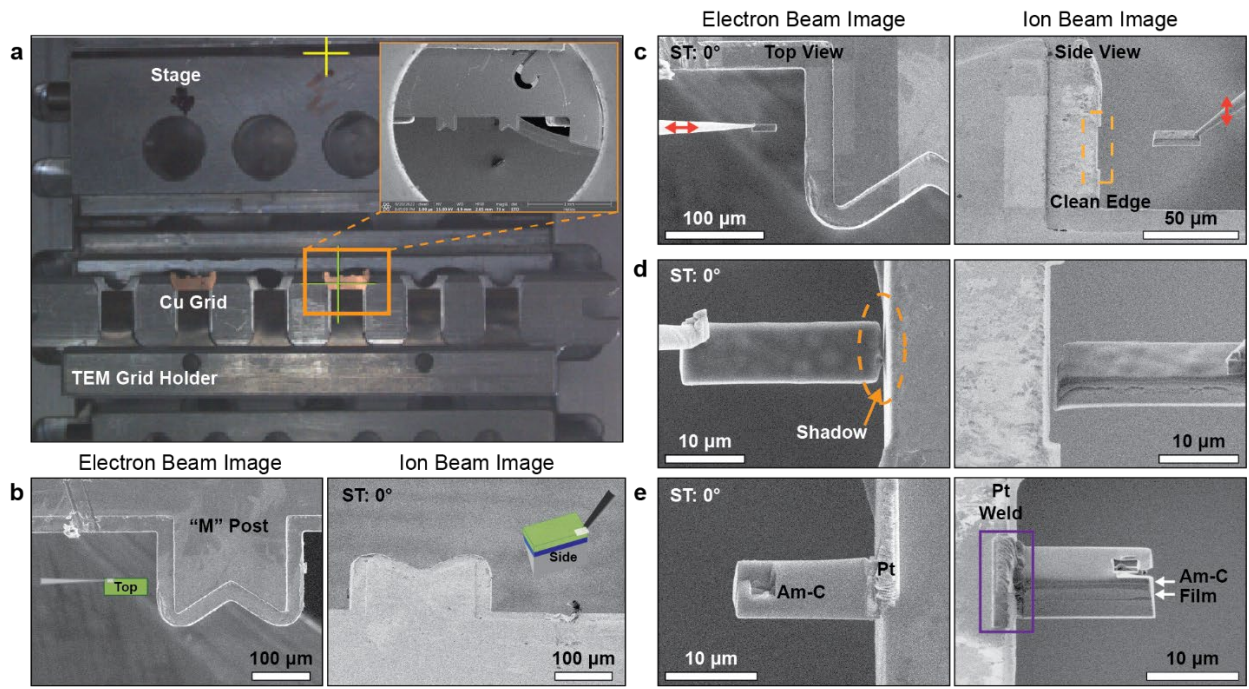

**Figure 3**: Sample attachment to Cu grid steps. (a) TEM grid holder for the Helios Dual Beam FIB mounted horizontally. Inset shows the high magnification image of one Cu Grid with two "M" posts available for sample mounting. (b) Electron-beam ("top") and Ion-Beam ("side") orientation of the "M" post edge used for sample mounting. (c) Easy-lift Needle with sample positioned on the grid at the eucentric stage position. The boxed area on the grid is trenched prior to create clean surface for sample attachment. (d) The needle is positioned close to the grid, and the contact can be observed by monitoring the "shadow" of the sample on the grid wall. Once contacted, the Pt needle is inserted to weld it to the Cu grid by depositing in the area marked in purple (e) The needle is cut off after the Pt deposition is complete, with a sample on the grid.

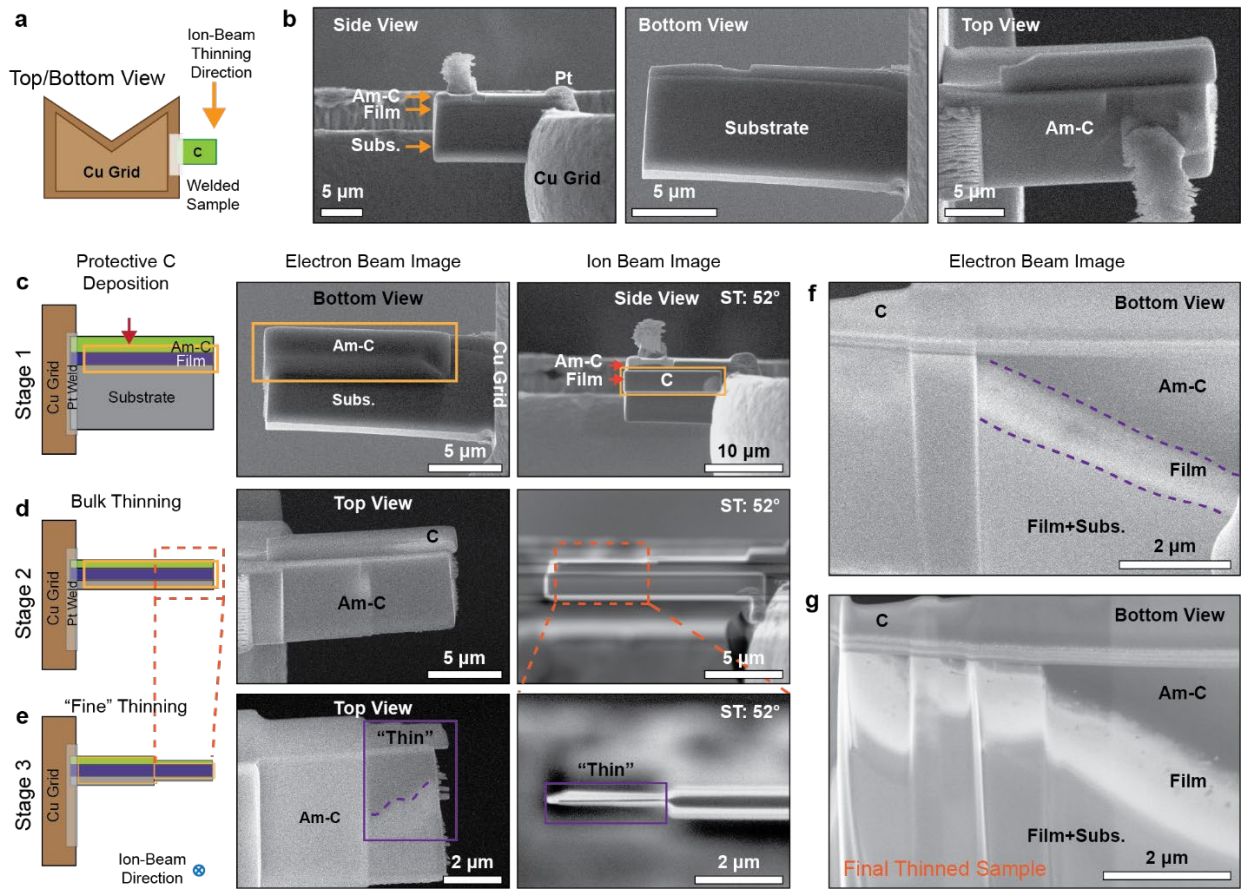

**Figure 4:** Sample thinning steps. (a) Model of sample attached to the Cu grid showing ion-beam thinning direction. (b) Electron-beam view at 0° tilt, showing the "wedge" in the side view. Electron-beam view of the bottom-side (substrate-side) and top-side (protective C side) at 52° stage tilt | 180° stage rotation. (c) Stage 1: Protective am-C is deposited in the ion-beam, covering the "film" area as marked by the orange box. (d) Stage 2: Higher beam current (0.44 nA) removal of bulk substrate from bottom/the protective carbon from the top. (e) Stage 3: Lower current thinning of substrate in windows. (f) SEM image from the "bottom"-side or the substrate-side at "5 keV" using the back-scatter mode, showing surface contrast between the film and the substrate. The bright contrast region enclosed by the purple lines are the "thin film" free of substrate. (g) Electron-beam image from final thinned sample in all areas.

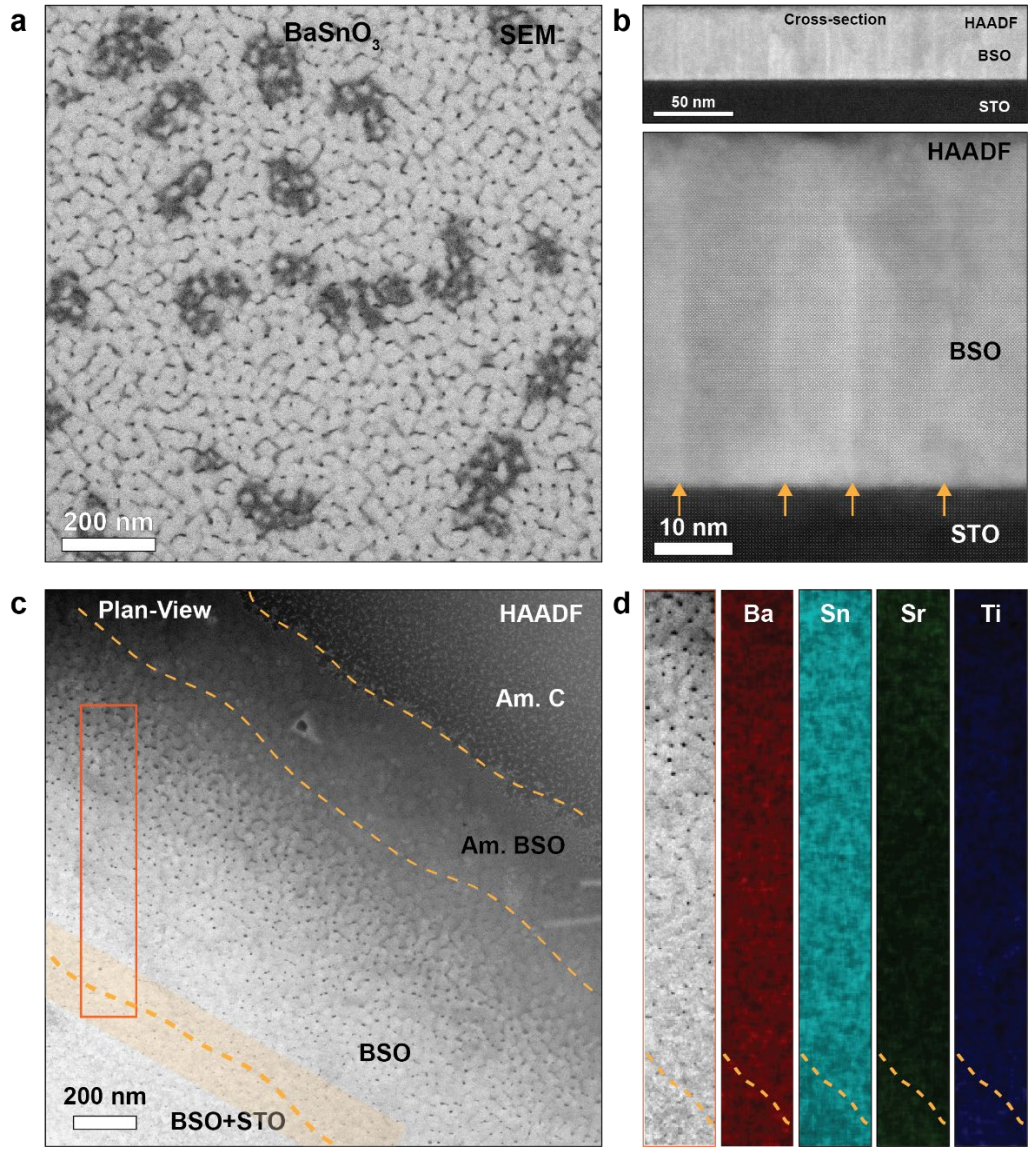

**Figure 5:** STEM analysis of BaSnO$_3$ films grown on SrTiO$_3$ substrate. (a) SEM image from the surface of a 50 nm BaSnO$_3$ (BSO) film grown on a SrTiO$_3$ (STO) substrate, showing the surface texture of the grains. Note: Areas of dark contrast are C contamination of the surface. (b) Cross-section HAADF image from the film, epitaxial growth is seen. Bright contrast lines marked by the yellow arrows correspond to defects. (c) HAADF-STEM image from a plan-view BaSnO$_3$ sample, showing areas of thin films with identical grain structure as seen in panel (a). (d) EDX maps from area highlighted in panel (c) showing most of the area of the film is free of the substrate SrTiO$_3$.

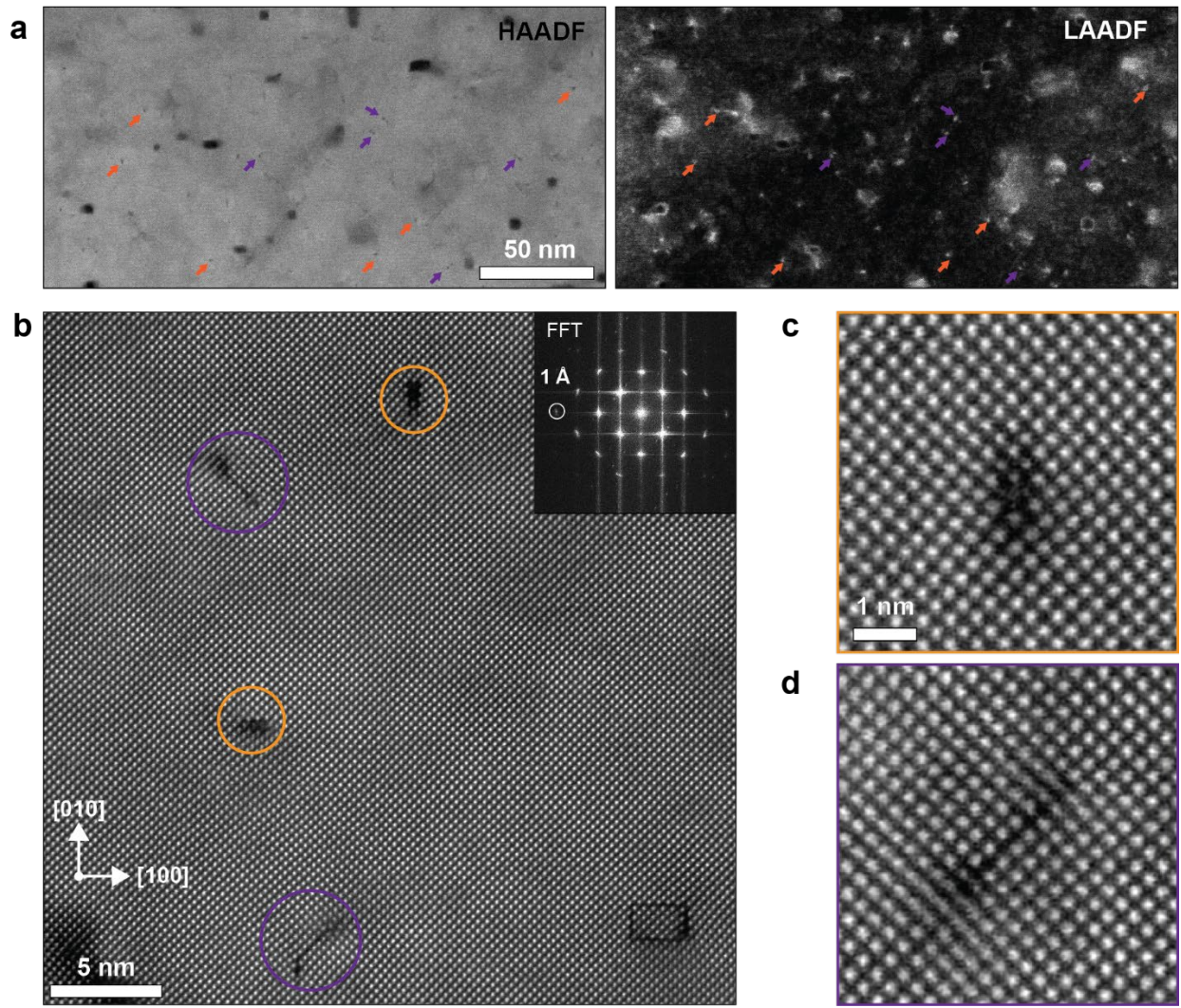

**Figure 6:** Identifying defects in BaSnO$_3$ films. (a) High magnification HAADF and low-angle (LA) ADF STEM images, where the dark and bright spots respectively correspond to defects. Single dislocation cores are marked by the orange arrows and dissociated dislocation cores are marked by purple arrows. (b) Atomic resolution HAADF image showing different dislocations in the thin film: single-cores ([001]/ (100)) highlighted with orange circles and dissociated-cores ([001]/ (110)) in purple. (c and d) Magnified atomic-resolution HAADF images of core structures of these single and dissociated dislocation core dislocations seen in BaSnO$_3$.

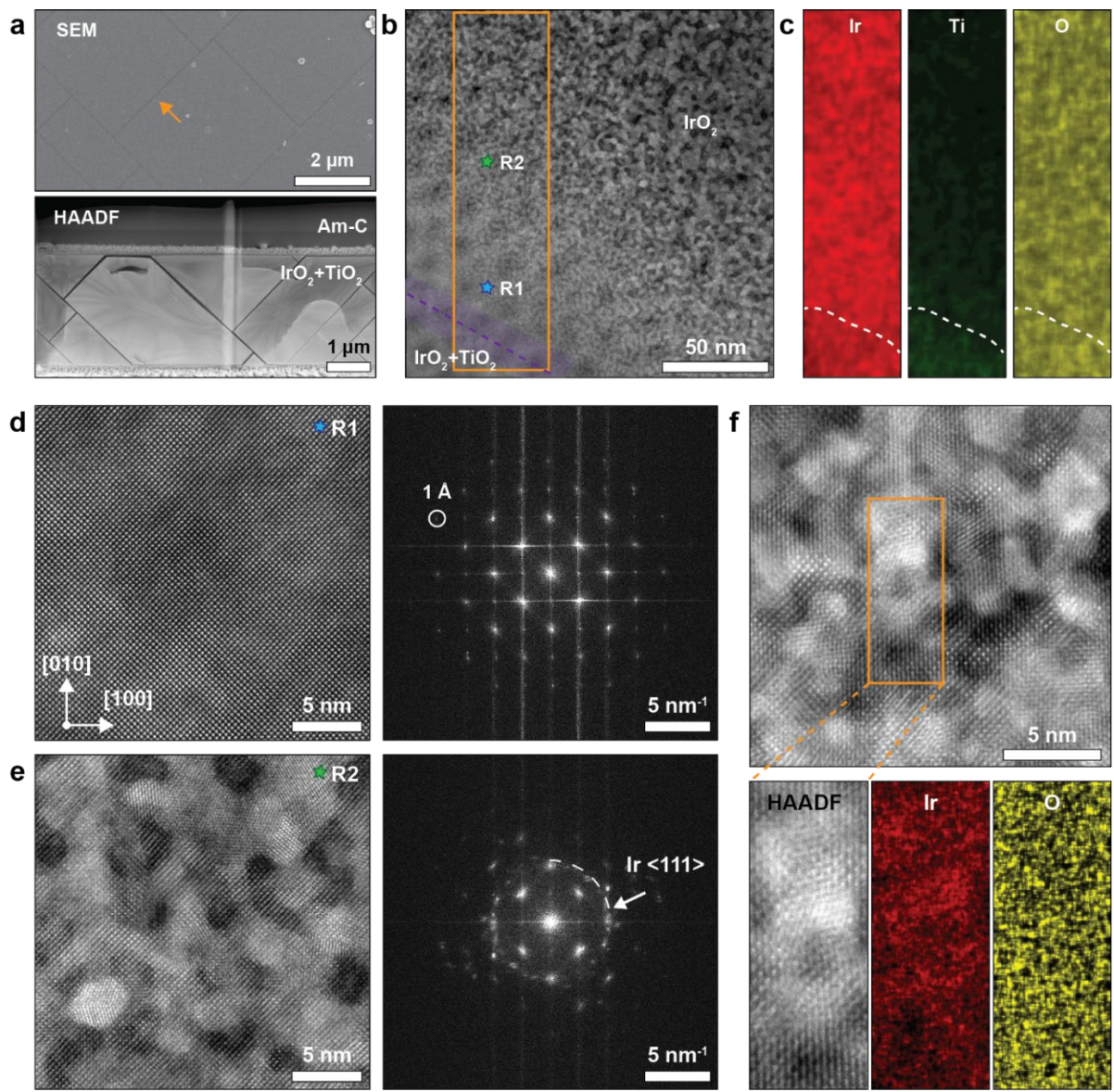

**Figure 7:** STEM analysis of IrO$_2$ films grown on TiO$_2$ substrate. (a) SEM image from film surface showing cracks on surface, with plan-view TEM sample from cracked area in bottom panel, where cracks are visible. (b) HAADF image from IrO$_2$ layer, showing varying film texture with thickness. (c) STEM-EDX maps from region highlighted in (b). (d) Atomic-resolution image from Region 1 (R1) marked in panel (b), showing IrO$_2$ along the (001) zone-axis, with the corresponding FFT. (e) Atomic resolution image from Region 2 (R1) marked in panel (b), showing texturing IrO$_2$, with the corresponding FFT showing ring corresponding to Ir <111> spacings of 2.26 Å. (f) Region highlighting Ir clusters on film surface with corresponding EDX maps showing higher Ir signals in nanoparticle areas.